\pdfoutput=1
\documentclass[%
aps,pra,twocolumn,showpacs,superscriptaddress,floatfix,longbibliography
]{revtex4-1}

\usepackage[dvipdfmx,dvipdfmx]{graphicx}
\usepackage{graphicx}
\usepackage{color}
\usepackage{siunitx}

\usepackage{subfigure}

\usepackage{epsfig}
\usepackage{float}
\usepackage{epstopdf}
\usepackage[dvipsnames]{xcolor}
\definecolor{myblue}{named}{Blue}
\definecolor{mygreen}{RGB}{0,120,0}
\usepackage[colorlinks=true,citecolor=myblue,linkcolor=myblue,urlcolor=myblue]{hyperref}

\usepackage{amsmath}
\usepackage{amsfonts}
\newcommand{\abs}[1]{\left| #1 \right|}
\newcommand{\ket}[1]{\left | #1 \right \rangle}
\def\k(#1){|#1\rangle}
\newcommand{\bra}[1]{\left \langle #1 \right |}
\newcommand{\proj}[1]{\ket{#1} \bra{#1}}

\newcommand{\beq}{\begin{equation}}
\newcommand{\eeq}{\end{equation}}
\newcommand{\beqa}{\begin{eqnarray}}
\newcommand{\eeqa}{\end{eqnarray}}
\newcommand{\beqan}{\begin{eqnarray*}}
\newcommand{\eeqan}{\end{eqnarray*}}

\newcommand{\affA}{%
\affiliation{
 Center for Macroscopic Quantum States (bigQ), Department of Physics, Technical University of Denmark, Building 307, Fysikvej, 2800 Kgs.~Lyngby, Denmark}
     }

\begin{document}

\title{
Adaptive generalized measurement for unambiguous state discrimination of quaternary phase-shift-keying coherent states}


\author{Shuro Izumi}
\email{sizumi@fysik.dtu.dk}
\affA
\author{Jonas S. Neergaard-Nielsen}
\affA
\author{Ulrik L. Andersen}
\affA

\begin{abstract}
Generalized quantum measurements identifying non-orthogonal states without ambiguity often play an indispensable role in various quantum applications.
For such unambiguous state discrimination scenario, we have a finite probability of obtaining inconclusive results and minimizing the probability of inconclusive results is of particular importance.
In this paper, we experimentally demonstrate an adaptive generalized measurement that can unambiguously discriminate the quaternary phase-shift-keying coherent states with a near-optimal performance.
Our scheme is composed of displacement operations, single photon detections and adaptive control of the displacements dependent on a history of photon detection outcomes.
Our experimental results show a clear improvement of both a probability of conclusive results and a ratio of erroneous decision caused by unavoidable experimental imperfections over conventional static generalized measurements.
\end{abstract}

\maketitle

\section{Introduction}\label{Sect:1}
Quantum measurements are ubiquitous in quantum information science.
Designing measurements to appropriately discriminate quantum states and efficiently reading out the information encoded in them is of general importance for quantum computation~\cite{Knill2001,RevModPhys.79.135}, sensing~\cite{Giovannetti1330,PhysRevLett.96.010401} and communication~\cite{Bennett,PhysRevLett.68.3121,RevModPhys.84.621}.
Quantum measurements in the framework of state discrimination can be classified into two major scenarios:
minimum error discrimination (MED) and unambiguous state discrimination (USD).
For MED, a measurement is designed to minimize the average error in discriminating quantum states~\cite{Helstrom_book76_QDET,Holevo_book}, 
whereas in USD, the aim is to discriminate quantum states with error-free conclusive decisions but with a finite probability of having inconclusive results~\cite{IVANOVIC1987257,DIEKS1988303,PERES198819}.
Measurements capable of unambiguously identifying non-orthogonal quantum states are desirable for applications where exact identification of the quantum states without ambiguity is required. Particular examples are quantum key distribution (QKD)~\cite{PhysRevA.51.1863,BANASZEK199912,PhysRevA.62.022306,PhysRevA.66.042313} and quantum digital signatures~\cite{gottesman2001quantum,Clarke2012, PhysRevLett.113.040502}.

The discrimination of coherent states is the principal example where non-trivial and carefully designed quantum measurements can enhance the performance over classical ones~\cite{Helstrom_book76_QDET}.
The MED of multiple coherent states has been well investigated, particularly for the binary phase-shift-keying (BPSK)~\cite{Kennedy73,Dolinar73,PhysRevLett.97.040502,PhysRevA.78.022320,PhysRevLett.101.210501,PhysRevLett.106.250503,PhysRevLett.121.023603} and the quaternary phase-shift-keying (QPSK) coherent state alphabets~\cite{Bondurant:s,PhysRevA.86.042328,PhysRevA.87.042328,Becerra2013_NP,Becerra2015_NP,PhysRevApplied.13.054015,PhysRevResearch.2.023384}.
It is widely acknowledged that displacement operations combined with photon detection provides a discrimination error that beats the shot noise limit achievable with conventional quadrature detection~\cite{PhysRevLett.101.210501,PhysRevLett.106.250503,PhysRevLett.121.023603}.
Furthermore, by including feedback control of the displacement operation, adapted by the detection events, it is possible to reach optimal MED performance for BPSK~\cite{Dolinar73} and near-optimal performance for QPSK signals~\cite{Bondurant:s,PhysRevA.86.042328,PhysRevA.87.042328} assuming that the feedback control is infinitely fast.
Although such an adaptive measurement scheme is technically challenging, its clear advantage has been experimentally observed in several experiments~\cite{Cook2007,Becerra2013_NP,Becerra2015_NP,PhysRevLett.124.070502,PhysRevApplied.13.054015,PhysRevResearch.2.023384}.
As for the USD strategy, generalized measurements enabling USD for $M$-ary PSK coherent states are realizable with $M$ displacement operations and photon detectors~\cite{PhysRevA.66.042313}.
This simple strategy without the complicated adaptive displacements is known to accomplish the optimal USD performance for BPSK coherent states: It maximizes the probability of identifying the states; in other words, minimizes the probability of inconclusive results~\cite{PhysRevLett.104.100505,PhysRevA.81.062338}.
On the other hand, in the more general case with multiple coherent states, $M>2$, there is a substantial gap between the optimal USD performance and the simple strategy~\cite{CHEFLES1998223,PhysRevA.66.042313,Becerra2013}. 

In this paper, we propose, theoretically investigate and experimentally demonstrate a generalized measurement scheme that unambiguously discriminates QPSK coherent states with a near-optimal success probability based on photon counting and real-time adaptive control of a displacement operation.
We show that, compared to conventional static protocols without adaptive control, our strategy is significantly enhancing the probability of successfully obtaining conclusive results while suppressing the ratio of erroneous decisions induced by unavoidable experimental imperfections. 

We first introduce our proposed generalized measurement with adaptive displacement operations in Sec.~\ref{Sect:2}.
In Sec.~\ref{Sect:3}, we present the results of our experimental demonstration and compare them with the conventional static schemes without adaptive control.
We conclude the paper in Sec.~\ref{Sect:4}.
\section{Adaptive generalized measurement for QPSK signals}\label{Sect:2}

\begin{figure}[t]
\centering 
{
\includegraphics[width=1.0\linewidth]
{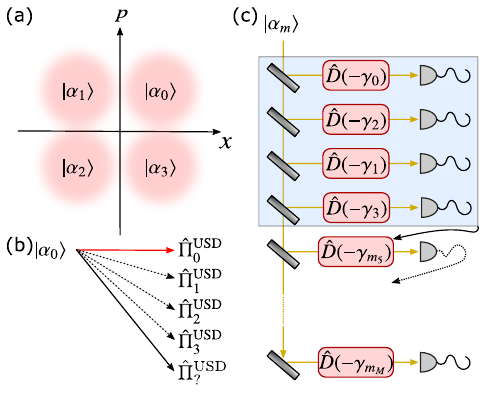}
}
\caption{
(a)
QPSK coherent states.
(b)
Transition diagram for signal $\ket{\alpha_0}$.
Red solid, black dashed and black solid lines correspond to conclusive and correct, conclusive but incorrect, and inconclusive results, respectively.
(c)
Schematic of USD measurement with displacement operations, single photon detectors and adaptive control of the displacement based on the photon detections. Shaded area indicates the static scheme with $M=4$. 
\label{theory_schematic}
}
\end{figure}

We consider the unambiguous discrimination of four coherent states defined as   
$\ket{\alpha_m}=\ket{\abs{\alpha} e^{(2m+1) i\pi/4}}$ where $m=0,\dots,3$ and $\abs{\alpha}$ represents the magnitude of the signal state. This set of coherent states is illustrated in phase space in Fig.~\ref{theory_schematic}(a).

A generalized measurement that allows to unambiguously discriminate the QPSK states can be represented by a positive operator valued measure (POVM) consisting of the elements $\{\hat{\Pi}_k^{\mathrm{USD}}, k=0,\dots,3\}$ for concluding $\ket{\alpha_k}$ and $\hat{\Pi}_{?}^{\mathrm{USD}}$ for inconclusive results.
These POVM elements satisfy the conditions
$\sum_{k=0}^3 \hat{\Pi}_{k}^{\mathrm{USD}}+\hat{\Pi}_{?}^{\mathrm{USD}}=\hat{I}$ and $\hat{\Pi}_{k}^{\mathrm{USD}},\hat{\Pi}_{?}^{\mathrm{USD}}\geq 0$.
A figure of merit of the USD is the total probability of obtaining conclusive results,
\begin{equation}
    P_C^{\mathrm{USD}}=1-\sum_{m=0}^3 p_m p_{?|m} = \sum_{m,k=0}^3 p_m p_{k|m},
\end{equation}
where $p_m$ is the {\it a priori} probability for $\ket{\alpha_m}$ which throughout this paper is assumed to be identical for the four states, $p_m=1/4$, and $p_{a|m} = \bra{\alpha_m} \hat{\Pi}_{a}^{\mathrm{USD}} \ket{\alpha_m}$ is the probability of obtaining 
the (potentially inconclusive) result $a$ given an incoming state $\ket{\alpha_m}$.
In Fig.~\ref{theory_schematic}(b),
we show a transition diagram for $\ket{\alpha_0}$.
The state is successfully identified without ambiguity if measurement results corresponding to the POVM element $\hat{\Pi}_0^{\mathrm{USD}}$ are obtained (the red arrow).
Some measurement results will be inconclusive. These are represented by the solid black arrow.
In an ideal USD measurement, the decisions $k=1,2,3$ represented by the dashed arrows will never occur, that is, $p_{k|m} = 0$ for $k \neq m$.
In practice, however, experimental imperfections will inevitably induce a certain amount of ambiguity in the measurement, resulting in non-zero $p_{k|m}$. The error probability will be our other figure of merit and is defined as the ratio of these erroneous conclusions to the total conclusive result probability:
\begin{equation}
    P_E^{\mathrm{USD}} = \frac{\sum_{m=0}^3 p_m \sum_{k\neq m} p_{k|m}}{P_C^{\mathrm{USD}}} .
\end{equation}

Our strategy for unambiguous discrimination of the non-orthogonal QPSK signals consists of beam splitters, displacement operations and single photon detectors (SPDs).
A schematic of this protocol is depicted in Fig.~\ref{theory_schematic}(c).
An input coherent state $\ket{\alpha_m}$ is equally split by the beam splitters into $M$ states $\ket{\gamma_m}$ with $\gamma_m=\alpha_m/\sqrt{M}$. Each of the states are then displaced and detected by an SPD. The displacement operations, $\hat{D}(-\gamma_i)=\exp(-\gamma_i \hat{a}^{\dagger} + \gamma_i^{\ast} \hat{a})$, are implemented such that one of the QPSK states is displaced to the vacuum state.
The SPD is capable of discriminating whether there exists at least one photon (``on'') or not (``off'').
These two outcomes are described by the POVM
$\{\hat{\Pi}^\mathrm{off}=e^{-\nu}
\sum_{n=0}^{\infty} (1-\eta)^n\ket{n}\bra{n}, \hat{\Pi}^\mathrm{on}=
\hat{I}-\hat{\Pi}^\mathrm{off}\}$, where $\nu$ is the dark count rate and $\eta$ is the detection efficiency, assumed to be the same for all SPDs.
Therefore, the probability of having an ``off'' outcome when the displacement $\hat{D}(-\gamma_i)$ is performed on a state $\ket{\gamma_m}$ is
\begin{align}
\label{eq:Poff}
    P(\mathrm{off}|m;i) &= \bra{\gamma_m}\hat{D}^{\dagger}(-\gamma_i)\hat{\Pi}^{\mathrm{off}}\hat{D}(-\gamma_i)\ket{\gamma_m} \\
     &= 
     \exp\left(-\nu-\eta\bar{n}_{m;i}\right) , \nonumber 
\end{align}
where $\bar{n}_{m;i}$ is the mean photon number of the signal state after displacement.
For ``on'' it is $P(\mathrm{on}|m;i) = 1 - P(\mathrm{off}|m;i)$.
Dark counts lead to false conclusive results since we may obtain an ``on'' event even when the incoming state is displaced to the vacuum state, i.e. $i=m$. 
The same may happen if the displacement operation on the beam-splitter has non-unity visibility. For a given visibility $\xi$, the mean photon number after displacement is
\begin{eqnarray}
\bar{n}_{m;i}&=&(1-\xi)(\abs{\gamma_m}^2+\abs{\gamma_i}^2)+\xi(\abs{\gamma_m-\gamma_i})^2
\nonumber
\\
&=&\frac{2\abs{\alpha}^2}{M}\Bigl(1-\xi\cos((m-i)\pi/2))\Bigr),
\end{eqnarray}
where the first and second terms in the first line represent the non-interfered and interfered components, respectively.

\begin{figure}[t]
\centering 
{
\includegraphics[width=1.0\linewidth]
{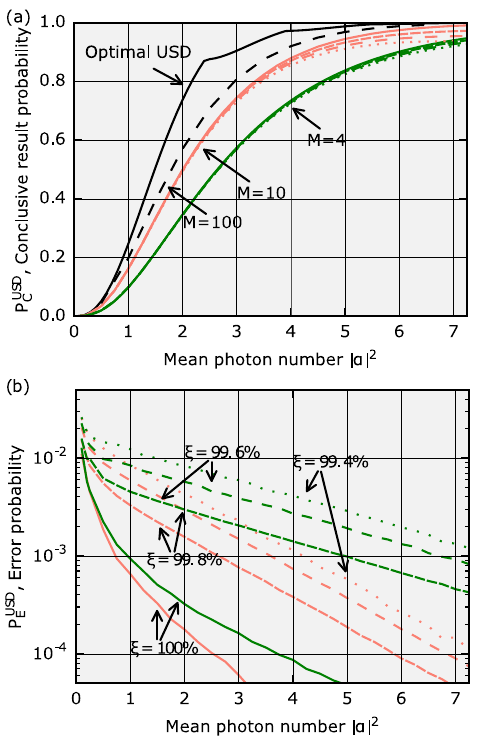}
}
\caption{
(a) Probabilities of conclusive results with unambiguous discrimination for QPSK signals with photon detection.
Black solid and dashed curves represent the fundamental bound of the probability of obtaining conclusive results for QPSK and the adaptive strategy with $M=100$ under ideal conditions with perfect visibility and no dark count, respectively.
Red curves are for $M=10$ and green curves for $M=4$.
The solid curves are for ideal conditions with perfect visibility and no dark count.
For more realistic conditions with experimental imperfections, the densely dashed, dashed and dotted curves are calculated for visibility conditions $\xi=99.8, 99.6$ and $99.4\%$, and fixed dark counts $\nu=1.0\times10^{-3}$.
(b) Probabilities of getting false conclusive results.
Solid, densely dashed, dashed and dotted lines indicate conclusive error probabilities with visibility condition $\xi=100, 99.8, 99.6$ and $99.4\%$, and fixed dark counts $\nu=1.0\times10^{-3}$. Again, green is for $M=4$, red for $M=10$.
\label{theory_success_error}
}
\end{figure}

A simple static USD strategy for the QPSK states can be performed with 4 stages ($M=4$), where the displacement operations are performed with four different phases. 
This strategy corresponds to the shaded area in Fig.~\ref{theory_schematic}(c).
Conclusive results are obtainable if any three of the SPDs give the outcome ``on''; otherwise the result is inconclusive.
The probability of having conclusive results for the USD measurement with $M=4$ can be analytically obtained from eq.~(\ref{eq:Poff}) and the decision scheme to be
\begin{eqnarray}
  P_C^{M=4} &=& \frac{1}{4}\sum_{m,k=0}^{3} p_{k|m} \nonumber \\
  &=& P_0(1-P_2)(1-P_1)^2 + P_2(1-P_0)(1-P_1)^2 \nonumber \\
  &&+\ 2P_1(1-P_1)(1-P_0)(1-P_2)
\end{eqnarray}
where $P_s = P(\mathrm{off}|m;i)$ when $i = m\pm s \text{ mod }4$, that is $P_s = e^{-\nu-(1-(1-s)\xi)\eta|\alpha|^2/2}$.
While this simple strategy enables us to unambiguously discriminate the QPSK coherent states (assuming no dark counts $\nu=0$ and perfect visibility $\xi=100\%$),  even for an ideal detection efficiency ($\eta=100\%$),
the probability of having such conclusive results is significantly lower than what is achievable by the theoretically optimal USD measurement (See Fig.~\ref{theory_success_error}(a) for $M=4$). This optimal performance can be derived by optimizing the POVM such that the conclusive result probability is maximized under the no error condition \cite{PhysRevA.66.042313,CHEFLES1998223}. In Appendix \ref{app:optimal_USD}, we outline this derivation for the QPSK signals.
 
We now turn to our proposed scheme for improving the probability of conclusive results. 
Here, we increase the number of splittings to some $M>4$ and maintain the static structure outlined above for the first four modes. For the remaining modes, though, the choice of displacement phase should now be dynamically adapted to the outcomes of all the previous SPD measurements. This introduces a temporal ordering of the modes which we can therefore consider as different stages of the receiver. See Fig.~\ref{theory_schematic}(c).
The displacement operation is set to displace the hypothetical state $\ket{\gamma_m}$ to the vacuum state in cyclic order of $m=0\rightarrow2\rightarrow1\rightarrow3\rightarrow0\rightarrow\cdots$. 
Detecting a photon eliminates the possibility of receiving the hypothetical state and the displacement is subsequently set to test for other hypothetical states in the following stages.
Hence, the displacement condition at the $j$'th stage is determined according to the counting history up to the $j-1$'th stage.
For example, the hypothetical state to be tested at the fifth stage, $m_5$, is set to be the first of the states in the cycle which hasn't yet been ruled out.
Finally, results are conclusive if and only if any three out of the $M$ SPDs detect photons.
Our adaptive strategy continues the displacement with a fixed hypothetical condition even after three ``on'' events are obtained and the results are regarded as inconclusive if 
a fourth ``on'' event occurs due to the visibility imperfection or the dark count.
It is worth noting that we may be able to improve the probability of conclusive results by making the final conclusion based on Bayesian inference of the received state
instead of regarding some cases, for example, as inconclusive results when having only four ``on'' events after obtaining a large number of ``off'' events with a fixed displacement condition. However, this would also increase the error probability since we conclude results with ambiguity. 
Furthermore, instead of changing the displacement in a cyclic manner, designing a more advanced adaptive strategy might lead to slightly improved performance.

We compare achievable probabilities of conclusive results in Fig.~\ref{theory_success_error}(a). 
Green and black solid curves represent the conclusive probabilities for the simple 4-stage strategy, with perfect visibility and no dark count, and the optimal ideal USD measurement \cite{CHEFLES1998223}, respectively.
Our strategy offers a major improvement to the static scheme, potentially closing most of the gap towards the optimal USD as evidenced by the red solid curve for $M=10$ and the black dashed curve for $M=100$ in a perfect visibility and no dark count condition. These probabilities are evaluated by Monte Carlo simulations.
Although the analysis of the performance in the asymptotic limit is not straightforward,
our numerical analysis indicates that the probability of conclusive results is saturated around 100 stages. 
We further evaluate the probabilities of conclusive results for various visibility conditions, $\xi=99.8$ (short dashed), $99.6$ (long dashed) and $99.4\%$ (dotted) with the finite dark count $\nu=1.0\times 10^{-3}$.
In the large mean photon regime, the performance of the adaptive strategy with $M=10$ is degraded because of the visibility imperfection while it is not critical for $M=4$.
Since we continue the measurement after having three ``on'' events and conclude the result as inconclusive if a fourth ``on'' event is obtained, 
the probability of having more than three ``on'' events increases, which reduces the probability of conclusive results.
A similar USD strategy relying on the adaptive displacement was discussed in \cite{PhysRevA.66.042313}, where an analytical expression of the conclusive probability was derived and the conclusive probability of its asymptotic limit $M\rightarrow\infty$ shows a similar performance with our scheme with $M=100$.

In Fig.~\ref{theory_success_error}(b), we show the error probabilities evaluated for the same values of $M$, $\xi$ and $\nu$ as in Fig.~\ref{theory_success_error}(a).
Evidently, the visibility imperfection is quite detrimental in terms of the error probability.
However, the adaptive strategy, while not only improving the conclusive result probability, can also significantly reduce the error, as shown here for $M=10$.
This reduction can be explained with similar reasoning as above: Instead of deciding on conclusive results after three ``on'' events, we continue to perform the displacement operation with a fixed hypothetical condition for the rest of the stages for confirmation. This reduces the probability of erroneously having conclusive results.
For very low mean photon numbers, the achievable error probability is mostly determined by the dark count rate.

Finally, we note that unambiguous discrimination can be emulated by heterodyne measurement if one is willing to accept the finite error of having conclusive results.
We will discuss the performance of the heterodyne measurement strategy and compare it with the performance of the photon detection protocol including experimental imperfections later in this paper.
The source code for the theoretical simulation of the strategies discussed in this paper is available at Ref.~\cite{torefersuppref}.
\section{Experiment}\label{Sect:3}

\begin{figure}[t]
\centering
{
\includegraphics[width=1.0 \linewidth]
{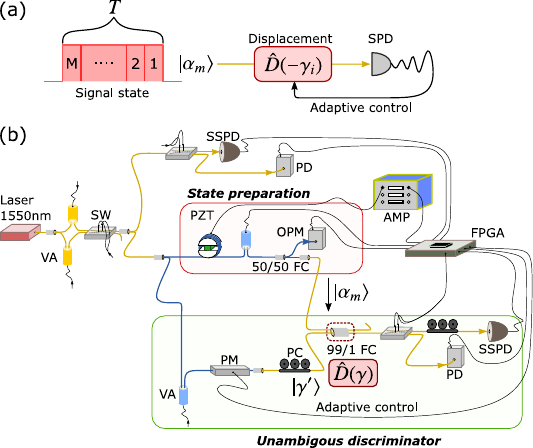}
}
\caption{
(a)
Schematic of temporal mode version of the adaptive strategy.
(b)
Experimental setup. Blue and yellow fibers respectively represent polarization maintaining fiber and single mode fiber.
FC, fiber coupler;
SW, optical switch;
PM, phase modulator;
PZT, piezo transducer;
VA, variable attenuator;
PC, polarization controller;
SSPD, superconducting nanowire single photon detector;
OPM, optical power meter;
AMP, amplifier.
}
\label{setup}
\end{figure}

In the previous section, we suggested to realize the adaptive measurement strategy by splitting the input state into multiple spatial modes. However, the measurement strategy can also be carried out in a temporal mode version where the state with the duration $T$ is virtually split into $M$ time bins and the displacement is successively updated in time through real-time feedback~\cite{PhysRevLett.97.040502}. For the experimental realization, we chose the temporal mode version as it only requires a single spatial mode, a single displacement operation and a single photon counter. A schematic of this version of the adaptive strategy is shown in Fig.~\ref{setup}(a).

\begin{figure}[t]
\centering
{
\includegraphics[width=1\linewidth]
{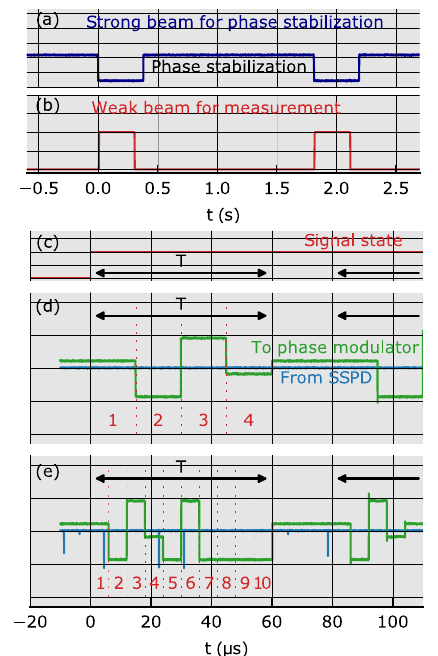}
}
\caption{
(a) Strong beam for phase stabilization. (b) Weak beam for measurement.
(c) Signal state having a time width $T=\SI{60}{\micro\second}$.
Signals applied to the phase modulator (green) and signals from the SSPD (blue) for (d) $M=4$ and (e) $M=10$.
\label{exp_procedure}
}
\end{figure}

Our experimental setup is shown in Fig.~\ref{setup}(b) and
we illustrate the sequencing of our experiment in Fig.~\ref{exp_procedure}.
A continuous-wave laser at 1550 nm (Koheras BASIK single frequency fiber laser model X15 from NKT Photonics) is split into two paths in order to prepare a signal state and a strong reference field for the displacement operation.
The intensity and the phase of the QPSK signal state are controlled with a variable attenuator and a phase shifter consisting of a piezo transducer embedded in a circular mount with an optical fiber looped around.
The signal state interferes with $1\%$ of the reference field $\ket{\beta^{\prime}}$ at a 99:1 fiber coupler, which enables us to implement the displacement operation.
By using an optical switch, the laser intensity is switched between high (Fig.~\ref{exp_procedure}(a)) and low (Fig.~\ref{exp_procedure}(b)) for the phase stabilization and the data acquisition, respectively.  The laser intensities for the strong and weak beams are individually stabilized by monitoring with a conventional photo detector and a superconducting nanowire single photon detector (SSPD) \cite{Miki:13,Yamashita:13}.
The relative phase between the signal and the reference is set to one of the four phase conditions $(2m+1)\pi/4$ $(m=0,\ldots,3)$ by monitoring an output of the 99:1 fiber coupler using a photo detector and directly locking to the correct points on the interference curve. Because of this locking method, the laser power stabilization is critical in order to avoid phase offsets in the displacement operations.
During the data acquisition period,
the output of the 99:1 fiber coupler is guided to an SSPD
and, instead of randomly preparing the QPSK coherent states, $300$ identical signal states are measured after releasing the phase stabilization loop (Fig.~\ref{exp_procedure}(c)).
This phase locking and data acquisition procedure is repeated 20 times for each of the QPSK states.
The voltage applied to the phase modulator, corresponding to the phase condition of the reference field, is controlled by a field programmable gate array (FPGA) dependent on the counting history of the SSPD. 
The FPGA implements the adaptive strategy using a hard-coded lookup-table.
Due to the finite response time (\SI{50}{\nano\second}) of the digital-to-analog converter that generates the signal to control the phase modulator, digital signal processing time in the FPGA to determine the output voltage (\SI{25}{\nano\second}) and electrical signal reflections, we discard a time interval of \SI{0.3}{\micro\second} between each time bin \cite{PhysRevApplied.13.054015}. This corresponds to a  $4.5(1.5)\%$ discarding loss for $M=10(4)$ 
since the temporal width of the signal state is defined to be $T=$ \SI{60}{\micro\second} in our experiment.

We achieve a total transmittance of $91\%$ from the 99:1 fiber coupler to the fiber right before the SSPD by splicing all optical fibers. The SSPD provides a detection efficiency around $73\%$ and a dark count of about \SI{27}{\hertz}.
Therefore, our system is able to accomplish a total system efficiency of about $66\%$.
For calibration of the system detection efficiency $\eta_{\mathrm{SE}}$,
a laser beam, propagating along the signal path, is divided into two paths, where one is used to monitor the laser power to estimate the power propagating along the other path.
The laser power is attenuated to the photon level by inserting a cascade of well calibrated optical fiber filters such that the SSPD can detect photons without being saturated.
The total system efficiency is evaluated by comparing the observed count rate $\eta_{\mathrm{SE}}\abs{\alpha}^2$ with the expected photon count rate $\abs{\alpha}^2$, where the latter is estimated from the laser power and the filter attenuation.
The total system detection efficiency can be estimated with around $1.5\%$ uncertainty including the finite precision of calibrating the filters, the uncertainty in the splitting ratio of the 50:50 fiber coupler and other systematic errors.

\begin{figure}[b]
\centering
{
\includegraphics[width=1\linewidth]
{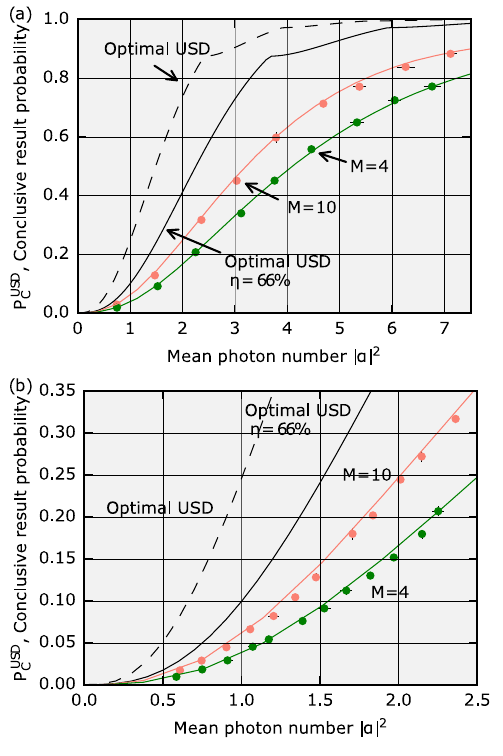}
}
\caption{
(a)
Probabilities of conclusive results with unambiguous discrimination for QPSK signals.
Circles indicate experimental results and
solid lines are theoretical predictions in the experimental condition.
Green is for $M=4$ and red is for $M=10$.
The theoretical upper bound for the conclusive result probability is shown by the black dashed line and the upper bound assuming a $66\%$ is represented by the black solid line.
(b)
Probability of conclusive results with unambiguous discrimination for QPSK signals with photon detections in the small mean photon number regime.
\label{exp_success}
}
\end{figure}

Examples of voltages applied to the phase modulator are shown in Figs.~\ref{exp_procedure}(d) and (e) respectively for $M=4$ and $M=10$.
The duration $T$ corresponding to the single signal state is divided into $4$ time bins in (d) and the voltage to the phase modulator shown by green is set to four different values to displace a signal state $\alpha_k$ to the vacuum in the order of $k=0\rightarrow2\rightarrow1\rightarrow3$.
Outputs from the SSPD are represented by blue. Since no photon is detected in (d), the final decision for this case is ``inconclusive''.
On the other hand, in Fig.~\ref{exp_procedure}(e), the voltage to the phase modulator is set to perform the displacement in the order of $k=0\rightarrow2\rightarrow1\rightarrow3$ for the first $4$ time bins and the following voltage condition is determined based on outcomes from the earlier time bins.
In this example, as the SSPD detects a photon in the first and fourth time bins and the possibilities of receiving the signal states $\alpha_0$ and $\alpha_3$ are eliminated, the voltage condition for the fifth time bin is set to displace $\alpha_2$ to the vacuum state. Furthermore, at the sixth time bin, detecting another photon indicates that the receiving signal state is not $\alpha_1$. Hence, the displacement is fixed to $\hat{D}(-\alpha_2/\sqrt{10})$ for the rest of the time bins (7---10) and we conclude that the receiving state is $\alpha_2$ because no photon is detected in the last $4$ time bins.

Experimental results for the probability of having conclusive results are shown in Figs.~\ref{exp_success}(a) and (b).
We evaluate the mean values and the error bars of the experimental results from 5 independent measurement runs. The error bars are negligibly small in most cases and estimated to be, at most, $8\%$. 
Since the conclusive result probability varies depending on the signal mean photon number, the power drift of our laser causes some errors on our experimental results as indicated by the error bars.
In some cases, relatively large noise from the environment affects our system 
which explains the inconsistency of the size of the error bars.
Green and red represent measurements for the static case of $M=4$ and the adaptive case for $M=10$, respectively.
The signal mean photon number $\abs{\alpha}^2$ is estimated by measuring the attenuated mean photon number, $\eta_{\mathrm{SE}}\abs{\alpha}^2$, directly by the SSPD, and then compensating for the losses.
We measure the attenuated mean photon number and the total detection efficiency multiple times before and after the data acquisition to evaluate the error bars on the mean photon number.
The theoretically attained maximum probability of conclusive results using the optimal USD strategy is shown by the black dashed line and the optimal USD with a $66\%$ loss is given by the black solid line.
The green and red lines represent the theoretical predictions for our measurement strategy assuming $\eta_{\mathrm{SE}}=66\%$, $\xi=99.4\%$, $\nu=1.5\times10^{-3}$ and the discarding loss because of the finite bandwidth of the digital-to-analog converter.
The experimental results show that our adaptive strategy is able to significantly enhance the probability of conclusive results. Moreover, the good agreement between the experimental results and the theory indicate that our system is well-controlled.

\begin{figure}[t]
\centering
{
\includegraphics[width=1\linewidth]
{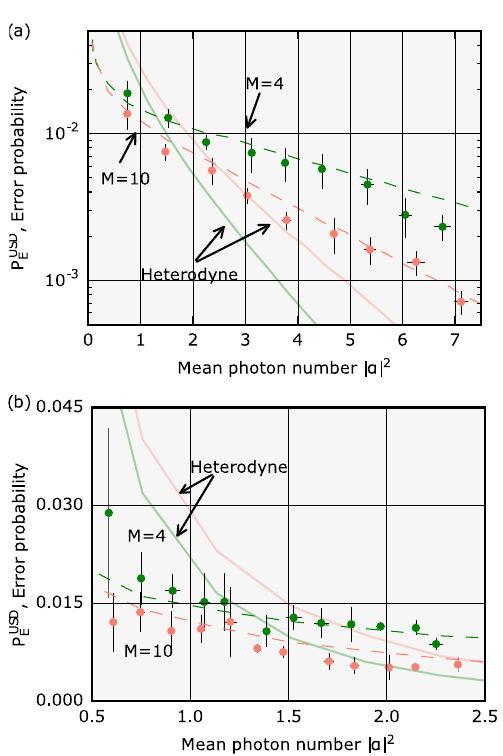}
}
\caption{
(a)
Probabilities of erroneously obtaining conclusive results.
Green and red circles are experimental results for $M=4$ and $M=10$, respectively.
Dashed lines are theoretical predictions in the experimental condition.
(b)
Probabilities of erroneously obtaining conclusive results in the small mean photon number regime.
Green and red solid lines represent the error probability given by heterodyne measurement when the heterodyne is designed to achieve the conclusive probabilities compatible with $M=4$ and $M=10$ shown by the solid lines in Fig.~\ref{exp_success}.
\label{exp_error}
}
\end{figure}

Furthermore, 
we evaluate error probabilities in Figs.~\ref{exp_error}(a) and (b).
Experimental results and theoretical predictions under the experimental condition are shown by circles and dashed lines.
A few discrepancies are observed between the theory and the experimental results.
These discrepancies, as well as the large error bars, are likely due to instability of the interference at the displacement operation. As can be seen from Fig.~\ref{theory_success_error}, even small changes in this parameter can have a big effect on the error probability.
Our setup has, without recalibration of the visibility, a variation of the visibility within $0.2\%$ due to polarization drift and unpredictable environmental noises.
The red and green lines in Fig.~\ref{exp_error} correspond to the predicted error probabilities when using a heterodyne measurement that is set to attain probabilities of conclusive results equal to those attained by the photon detection scheme with $M=10$ and $M=4$ corresponding to the solid lines in Fig.~\ref{exp_success}.
Throughout the paper, we calculate the performance of the ideal ($100\%$ efficient) heterodyne measurement for the comparison to our experimental results.
For the heterodyne measurement, the USD is emulated by applying thresholding in phase space and subsequently 
post processing the obtained outcomes. The POVM for inconclusive results can be represented as $\hat{\Pi}_{?}=\int_{{x,p}\in\mathcal{S}}\ket{x}\bra{x}\otimes\ket{p}\bra{p}dxdp$ for some region $\mathcal{S}$ of phase space.
There exists a trade-off between the probabilities associated with errors and conclusive results and therefore the post processing is optimized such that the error probability is minimized while achieving a certain probability for a conclusive result. We adapt a linear thresholding approach for simplicity, where the region is chosen to be
$\mathcal{S}= \{(x,p) |\ |x| < x_{th} \lor |p| < p_{th}\}$ and $x_{th}=p_{th}$.
More advanced post processing can be performed by numerically optimizing the region $\mathcal{S}$, but the improvement over the linear thresholding strategy is very small \cite{Sych_2010,Becerra2013}. 
Further mathematical description regarding the implementation of the USD with the heterodyne measurement is given in Appendix \ref{app:hetero_USD}.
Our adaptive measurement strategy shows a clear improvement of the error probability over the conventional simple protocol with $M=4$ for all $|\alpha|^2$, and moreover, it beats the heterodyne strategy in the low photon number regime up to approximately $|\alpha|^2=2.5$.
Beating the heterodyne strategy in the higher photon number regime is also possible by installing state-of-the-art photon detectors with a high detection efficiency \cite{Marsili2013}. A further discussion for the performance in various detector efficiency conditions is provided in Appendix \ref{app:EAVDEC}.

\begin{figure}[t]
\centering
{
\includegraphics[width=1\linewidth]
{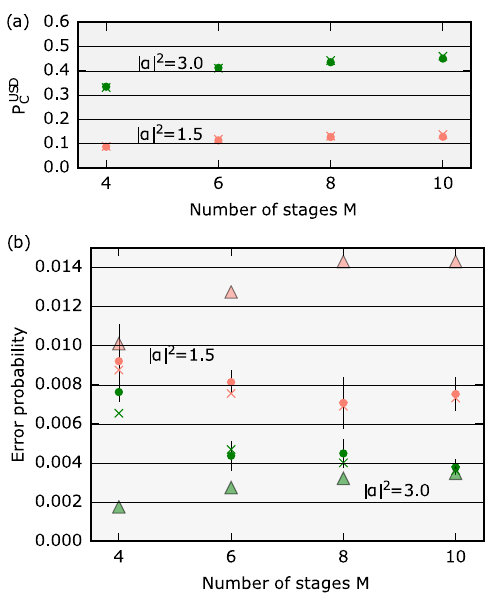}
}
\caption{
(a)
Probabilities of conclusive results for unambiguous discrimination of QPSK states as a function of the number of stages.
Circles and crosses represent experimental results and theoretical predictions, respectively.
Red and green circles are the experimental results for $\abs{\alpha}^2=1.5$ and $\abs{\alpha}^2=3.0$.
(b)
Error probabilities of having conclusive results.
Circles and crosses are experimental results and theoretical prediction, and triangles are error probabilities for heterodyne measurement that is designed such that it achieves a conclusive probability corresponding to the performance of the photon detection scheme shown in (a).
Red and green are respectively for $\abs{\alpha}^2=1.5$ and $\abs{\alpha}^2=3.0$.
\label{results_NumberM}
}
\end{figure}

To demonstrate the improvement in performance when increasing the number of detection stages, we show in Fig.~\ref{results_NumberM} the results of the conclusive (a) and the error (b) probabilities for detectors with varying number of stages. We performed the measurements for two different fixed mean photon numbers, $\abs{\alpha}^2=1.5$ (red) and $\abs{\alpha}^2=3.0$ (green).
Theoretical predictions under the given experimental condition with $\eta=66\%,\xi=99.55\%, \nu=1.5\times10^{-3}$ as well as the discarding loss for delay compensation, are represented by crosses.
Triangles correspond to the error probabilities for the heterodyne measurement under the condition that the probabilities of conclusive results achieve the experimentally obtained probabilities shown by the circles in Fig.~\ref{results_NumberM}(a).
Red and green colors are associated with the mean photon number $\abs{\alpha}^2=1.5$ and $\abs{\alpha}^2=3.0$, respectively.
Both the probabilities of conclusive results and of errors can be improved by increasing the number of stages but the improvement saturates (or even slightly degrades) for a large number of stages since the discarding loss becomes dominant as the number of stages increases.
Finally, we expect that our adaptive strategy is not an unbiased measurement, i.e., the conclusive probabilities and the error probabilities for each signal condition are not uniform while the static scheme with $M=4$ is, in the ideal case, unbiased.  We provide a further analysis of the bias induced in our experiment due to its design and experimental imperfections in Appendix \ref{app:biasanalysis}.

\section{Conclusions}\label{Sect:4}
We proposed and experimentally realized an adaptive generalized quantum measurement that unambiguously discriminate quaternary phase-shift-keying coherent states with a near-optimal performance.
Our strategy consists of a displacement operation, a single photon detector and real-time adaptive control of the phase space displacements that depend on the history of single photon detection outcomes.
We demonstrated the adaptive generalized measurement and, while the performance is degraded due to the finite efficiency of our system, we observed a clear improvement of the probability of having conclusive results in comparison with a simple scheme using a non-adaptive approach.
Furthermore, 
we evaluated the probability of erroneously obtaining conclusive results which is caused by the non-perfect interference contrast of the displacement operation as well as the dark counts.
By increasing the number of detection stages, the error probability can be suppressed, yielding better performance than a heterodyne measurement designed to reach a comparable probability of conclusive results for a wide range of signal mean photon numbers.  

Since adaptive phase space displacements based on photon detections provide near-optimal performance for the minimum error discrimination of multiple coherent states in addition to the unambiguous state discrimination,
it is expected to serve as a novel receiver technique in applications associated with classical coherent communication \cite{7174950} as well as quantum communication, in particular quantum key distribution  \cite{pirandola2019advances}.
To install such a receiver in a real communication scenario, the preparation of a stable reference field at the receiver side for displacements would be required. 
This can be done by transmitting along with the signal a coherent field extracted from the same laser, either in a separate fiber or multiplexed with the signal. At the receiver side, this reference field can either be used directly for the displacement (similar to what we did here) or for synchronizing the phase and frequency of a locally generated displacement field \cite{pirandola2019advances}.
By establishing these techniques,
we believe that the adaptive phase space displacements based on photon detections will be a beneficial tool for real applications beyond our proof-of-concept experiment.

\begin{acknowledgments}

This project was supported by Grant-in-Aid for JSPS Research Fellow, by VILLUM FONDEN via the Young Investigator Programme (Grant no. 10119), by EU project CiViQ (Grant no. 820466) and by the Danish National Research Foundation through the Center for Macroscopic Quantum States (bigQ DNRF142). 
S.I and U.L.A acknowledge support from the EU Horizon 2020 Research and Innovation Program  under Grant Agreement No. 862644 (Quantum Readout Techniques and Technologies, QUARTET).
S.I and J.S.N.-N designed and developed the experimental setup.
S.I conducted the experiment and analyzed the experimental data.
S.I, J.S.N.-N and U.L.A conceived the original idea and wrote the manuscript.
\end{acknowledgments}

\appendix
%
\section{Optimal USD performance} \label{app:optimal_USD}
For a set of $M$ signals $\{\ket{\psi_m}\}$ satisfying a specific symmetric condition (such as the even distribution in phase space of $M$-PSK coherent states), 
the maximum achievable conclusive probability for the USD was found by Chefles and Barnett \cite{CHEFLES1998223}:
\begin{equation}
P_{\mathrm{USD}}=M\min_{k} \abs{c_k}^2.
\end{equation}
Here, $\{c_k\}$ are the coefficients that appear when introducing an appropriate set of orthogonal vectors $\{\ket{\omega_k}\}$ whose linear combination constitutes the original symmetric signal states in the specific form
\begin{equation}
\ket{\psi_m}=\sum_{k=0}^{M-1} c_k u^{mk} \ket{\omega_k},
\end{equation}
where $u=e^{\frac{2\pi i}{M}}$.
For $M$-PSK coherent signals, the general expression for $\abs{c_k}^2$ is found to be \cite{PhysRevA.66.042313,CHEFLES1998223}, 
\begin{equation}
\abs{c_k}^2=\frac{1}{M}\sum_{j=0}^{M-1} e^{-2\pi i jk/M}e^{\abs{\alpha}^2(e^{2\pi ij/M}-1)},
\end{equation}
which for QPSK signals become
\begin{eqnarray}
\abs{c_0}^2&=&\frac{e^{-\abs{\alpha}^2}}{2} (\cosh{\abs{\alpha}^2}+\cos{\abs{\alpha}^2}),
\nonumber
\\
\abs{c_1}^2&=&\frac{e^{-\abs{\alpha}^2}}{2} (\sinh{\abs{\alpha}^2}+\sin{\abs{\alpha}^2}),
\nonumber
\\
\abs{c_2}^2&=&\frac{e^{-\abs{\alpha}^2}}{2} (\cosh{\abs{\alpha}^2}-\cos{\abs{\alpha}^2}),
\nonumber
\\
\abs{c_3}^2&=&\frac{e^{-\abs{\alpha}^2}}{2} (\sinh{\abs{\alpha}^2}-\sin{\abs{\alpha}^2}).
\end{eqnarray}
While each $\abs{c_k}$ is continuous and smooth as a function of the signal mean photon number $\abs{\alpha}^2$, the minimum value varies depending on $\abs{\alpha}^2$. This results in the step-like behavior of the performance as one can see from Fig.~\ref{theory_success_error}.

\section{Performance analysis of heterodyne measurement}\label{app:hetero_USD}

To perform the USD with the heterodyne measurement, the POVM of the heterodyne measurement for conclusive results can be represented by 
\begin{equation}
\hat{\Pi}_m=\int_{x,p\in \mathcal{S}_m} \proj{x}\otimes\proj{p} dxdp,
\end{equation}
where $\mathcal{S}_m$ ($m=0,1,2,3$) is a region of phase space for which the outcome is concluded as state $m$.
The POVM for inconclusive results is 
\begin{equation}
\hat{\Pi}_?=\hat{I}-\sum_{m=0}^3\hat{\Pi}_m=\int_{x,p\in \mathcal{S}_i} \proj{x}\otimes\proj{p} dxdp,
\end{equation}
where the results are inconclusive if the outcomes from the heterodyne measurement belong to a region $\mathcal{S}_i$ of phase space.\
In contrast to a measurement strategy based on displacements and single photon detectors, 
the USD with heterodyning cannot perform the ideal USD, i.e., zero error probability is not achievable.
While the inconclusive probability can be decreased if one sets the region $\mathcal{S}_i$ small, the error probability,  
$\bra{\alpha_k}\hat{\Pi}_m \ket{\alpha_k}\neq 0$ for $m\neq k$, becomes large.
Therefore, there is a trade off relation for heterodyning between the conclusive result probability and the error probability.
To analyze the performance of the heterodyne measurement and compare it with our strategy, we first set the desired conclusive result probability and determine the regions $\mathcal{S}_m$ and $\mathcal{S}_i$.
Then the error probability for the heterodyne that is designed to achieve the desired conclusive result probability can be evaluated from these regions.
The conclusive result probability and the error probability for the heterodyne can be represented as,
\begin{eqnarray}
P_c^H &=&1- \frac{1}{4}\sum_{m=0}^3\bra{\alpha_m} \hat{\Pi}_? \ket{\alpha_m},
\\
P_e^H &=& \frac{1}{4}\frac{\sum_{m=0}^3 \sum_{k\neq m} \bra{\alpha_m} \hat{\Pi}_k \ket{\alpha_m}}{P_c^H},
\end{eqnarray}
where $1/4$ originates from the equal a priori probabilities.
To evaluate the best performance accomplished by the heterodyne, the regions need to be optimized \cite{Sych_2010,Becerra2013}.
In our analysis, instead of optimizing the regions, we design the POVMs of the heterodyne detection as,
\begin{eqnarray}
\hat{\Pi}_0&=&\int_{x_{th}}^{\infty}\int_{p_{th}}^{\infty} \proj{x}\otimes\proj{p} dxdp,
\nonumber
\\
\hat{\Pi}_1&=&\int_{-\infty}^{-x_{th}}\int_{p_{th}}^{\infty} \proj{x}\otimes\proj{p} dxdp,
\nonumber
\\
\hat{\Pi}_2&=&\int_{-\infty}^{-x_{th}}\int_{-\infty}^{-p_{th}} \proj{x}\otimes\proj{p} dxdp,
\nonumber
\\
\hat{\Pi}_3&=&\int_{x_{th}}^{\infty}\int_{-\infty}^{-p_{th}} \proj{x}\otimes\proj{p} dxdp,
\nonumber
\\
\hat{\Pi}_?&=&\hat{I}-\sum_{i=0}^3 \hat{\Pi}_i=\int_{x,p\in \mathcal{S}_i} \proj{x}\otimes\proj{p} dxdp, \nonumber
\\
\mathcal{S}_i&=&\{(x,p) | |x| < x_{th} \lor |p| < p_{th}\},
\end{eqnarray}
and choose the thresholds $z_{th}=x_{th}=p_{th}$. For this strategy, the conclusive result probability and the error probability of the heterodyne measurement can be expressed in the analytical form, 
\begin{widetext}
\begin{eqnarray}
P_c^H &=&\frac{1}{4}\Bigl\{\mathrm{erfc}(z_{th}-\sqrt{n/2})+\mathrm{erfc}(z_{th}+\sqrt{n/2})\Bigr\}^2,
\\
P_e^H &=& \frac{1}{4} \frac{\mathrm{erfc}(z_{th}+\sqrt{n/2})\Bigl\{\mathrm{erfc}(z_{th}+\sqrt{n/2})+2\mathrm{erfc}(z_{th}-\sqrt{n/2})\Bigr\}}{P_c^H},
\end{eqnarray}
\end{widetext}
where $\mathrm{erfc}(x)$ is the complementary error function defined as
$\mathrm{erfc}(x)=\frac{2}{\sqrt{\pi}}\int_x^{\infty} \mathrm{e}^{-t^2} dt$ and $n$ is the signal mean photon number.

\section{Error analysis in various detector efficiency conditions}\label{app:EAVDEC}
\begin{figure}[t]
\centering
{
\includegraphics[width=1\linewidth]
{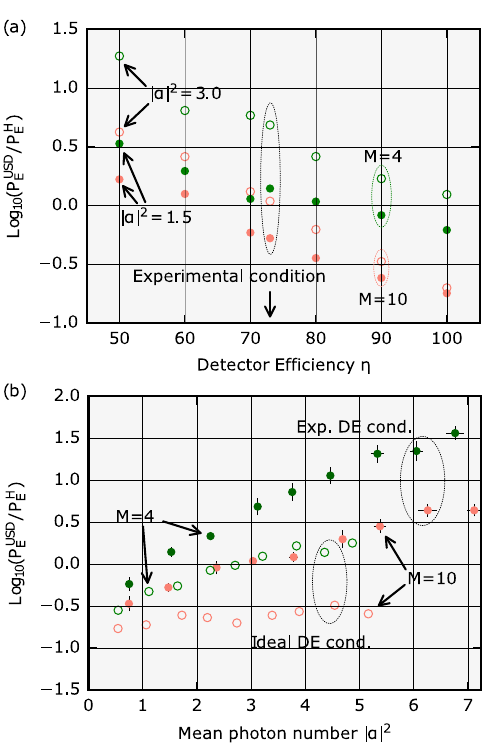}
}
\caption{
(a) Ratio of the error probabilities for our strategy with various detection efficiencies and the heterodyne measurement.
Open and filled circles represent the data for $\abs{\alpha}^2=1.5$ and $\abs{\alpha}^2=3.0$.
Green and red circles are for $M=4$ and $M=10$.
(b) Ratio of the error probabilities for our strategy and the heterodyne measurement.
Filled and open circles are the ratio in experimental detector efficiency condition (73$\%$) and ideal detector efficiency condition (100$\%$), respectively.
Green and red circles are for $M=4$ and $M=10$.
\label{exp_error_VSHetero}
}
\end{figure}

%
By performing the adaptive strategy, we observe the improvement of the probability of conclusive results while suppressing the error probability due to the dark count and visibility imperfection.
However, 
a heterodyne measurement designed to approach the comparable conclusive results probability yields a better error probability in the large mean photon number regime. 
Here, we analyze the error probability achievable by our strategy with an SPD under various detection efficiency conditions and compare them with the ideal heterodyne detection strategy with a $100\%$ system efficiency.
Since our system has a finite transmittance of $91\%$ from the signal preparation point to the SSPD, 
the maximum achievable total system efficiency is limited to $91\%$.
In Fig.~\ref{exp_error_VSHetero}(a),
we plot the ratio of the error probabilities of our strategy and the heterodyne measurement as a function of the quantum efficiency of the SPD for the mean photon numbers $\abs{\alpha}^2=1.5$ (filled circles) and $\abs{\alpha}^2=3.0$ (open circles).
Green and red colors indicate the results for $M=4$ and $M=10$, respectively.
The actual experimental condition corresponds to the detector efficiency of $73\%$, which gives a total system efficiency of $66\%$.
To obtain Fig.~\ref{exp_error_VSHetero}(a),
we first normalize the experimentally measured mean photon number with the detector efficiency of our SSPD ($73\%$) and multiply by the detector efficiency of the imaginary SPD, and then choose the mean photon number closest to the $\abs{\alpha}^2=1.5$ and $\abs{\alpha}^2=3.0$ from a list of re-scaled mean photon numbers.
Using the mean photon number and corresponding probability of conclusive results, we calculate the error probability for the heterodyne detection strategy in each detector efficiency condition.
For $\abs{\alpha}^2=1.5$, a detector efficiency of more than $70\%$ is sufficient to beat the heterodyne limit by using the adaptive strategy with $M=10$.
On the other hand, for the simple static strategy with $M=4$, a detection efficiency of more than $90\%$ is required to surpass the heterodyne limit.
For $\abs{\alpha}^2=3.0$, the static protocol cannot beat the heterodyne limit even if we employ an ideal SPD with unit quantum efficiency.
On the other hand, by using a state-of-the-art SPD with a detector efficiency of more than $80\%$, our adaptive detection strategy is able to outperform the ideal heterodyne detector. 

Furthermore, we investigate the ratio of the error probabilities as a function of the mean photon number, both using the actual experimental condition with a detector efficiency of 73$\%$ (filled circles) and the ideal condition of a unit detector efficiency (open circles). These results are shown in Fig.~\ref{exp_error_VSHetero}(b).
Our analysis indicates that our adaptive measurement strategy provides an error probability that is lower than the error probability of the heterodyne detector in a wide range of mean photon numbers if an SPD with high efficiency is available. On the other hand, it is clear that the simple static strategy is not able to beat the heterodyne measurement limit for large mean photon numbers.

\section{Bias analysis of adaptive strategy}\label{app:biasanalysis}
\begin{figure}[h]
\centering
{
\includegraphics[width=1\linewidth]
{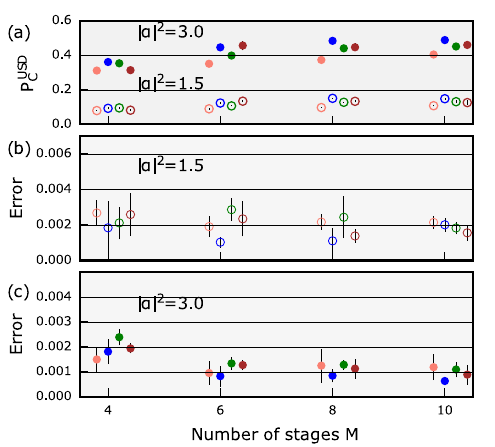}
}
\caption{
(a)
Probabilities of conclusive results for each signal state as a function of the number of stages.
Open and filled circles represent experimental results for $\abs{\alpha}^2=1.5$ and $\abs{\alpha}^2=3.0$, respectively.
Red, green, blue and brown represent the experimental results for $m=0$,$1$,$2$ and $3$.
Error probabilities of having conclusive results for (b) $\abs{\alpha}^2=1.5$ and (c) $\abs{\alpha}^2=3.0$.
\label{error_NumberM_individual}
}
\end{figure}

The simple strategy consisting of four displacements without adaptive control ($M=4$) is an unbiased measurement, i.e., the conclusive result probability as well as the error probability is the same for all signal conditions.
On the other hand, since we perform the displacement operation such that a state $\alpha_m$ is displaced to the vacuum state in the order of $0\rightarrow2\rightarrow1\rightarrow3$, the adaptive strategy with $M>4$ exhibits a finite bias dependent on the incoming signal state condition.
In Fig.~\ref{error_NumberM_individual}, we represent the conclusive result probabilities and the error probabilities for each signal condition individually as a function of the number of stages $M$. Red, green, blue and brown represent the experimental results for $m=0$,$1$,$2$ and $3$.
We show the conclusive results probabilities for $\abs{\alpha}^2=1.5$ and $\abs{\alpha}^2=3.0$ with open and filled circles respectively in Fig.~\ref{error_NumberM_individual} (a).
The mean and the error bars are evaluated from 5 independent procedures and the error bars are negligibly small. 
As we can see from Fig.~\ref{error_NumberM_individual} (a), the conclusive result probabilities for the measurement with $M=4$ show very similar values for all signal conditions. 
As opposed to the static strategy with $M=4$, for the adaptive strategy with $M>4$, there are finite biases on the conclusive result probabilities dependent on the signal state condition; the conclusive result probability obtained for $\alpha_0$ tends to be smaller than others.
This is due to the design of the adaptive strategy. As we perform the displacement in the order of $0\rightarrow2\rightarrow1\rightarrow3$, we have a higher probability of successfully eliminating the signal state $\alpha_0$.  
With respect to the error probabilities shown in Fig.~\ref{error_NumberM_individual} (b) and (c) respectively for $\abs{\alpha}^2=1.5$ and $\abs{\alpha}^2=3.0$, the dependence on the incoming signal condition is not as consistent as that of the conclusive result probability because the error probability is rather sensitive to the variation of the visibility and the calibration of the displacement phase.

\bibliography{reference}

\end{document}